%% file: main.tex
\documentclass[preprint,3p,times,onecolumn, sort&compress]{elsarticle}
\pdfoutput=1
\usepackage{etoolbox}
\usepackage[utf8]{inputenc}
\usepackage[colorlinks]{hyperref}
\usepackage{graphicx}
\usepackage{tabularx}
\usepackage{lipsum}
\usepackage{wrapfig}
\usepackage{amsmath}
\usepackage{float}
\usepackage{subfigure}
\usepackage{gensymb}
\usepackage{hyperref}
\usepackage{cleveref}
\usepackage{chngcntr}
\usepackage{multirow}
\usepackage{makecell}
\usepackage{array}
\usepackage{booktabs}
\usepackage[none]{hyphenat} %% no words with hyphens
\usepackage{enumitem}
\usepackage{ragged2e}
\usepackage{amssymb}
\usepackage[section]{placeins}
\usepackage{siunitx}
\usepackage{epigraph}
\usepackage{lineno}
\usepackage{braket}
\usepackage{xparse,nameref}
\usepackage[usestackEOL]{stackengine}
\usepackage{xcolor}
\usepackage{mathtools}
\usepackage{multirow}
\usepackage{makecell}
\usepackage{array}
\usepackage{booktabs}
\usepackage{textcase}
\usepackage[toc,page]{appendix}
\usepackage{pdfpages}
\usepackage{titlesec}
\usepackage{natbib}

\titleformat{\subsection} {\normalfont\bfseries}{\thesubsection}{1em}{}
\titleformat{\paragraph} {\normalfont\bfseries\itshape}{\paragraph}{1em}{}
\usepackage{dcolumn}   % needed for some tables
\usepackage[tablename=TABLE, labelsep=period, figurename=Figure, skip=0pt, font=small]{caption}

\makeatletter
\def\ps@pprintTitle{%
 \let\@oddhead\@empty
 \let\@evenhead\@empty
 \def\@oddfoot{}%
 \let\@evenfoot\@oddfoot}
\makeatother

\begin{document}
  \begin{frontmatter}
       \title{\textbf{First Directional Measurement of sub-MeV Solar Neutrinos with Borexino}}
\author[London1,Munchen]{M.~Agostini}
\author[Munchen]{K.~Altenm\"{u}ller}
\author[Munchen]{S.~Appel}
\author[Kurchatov]{V.~Atroshchenko}
\author[Juelich]{Z.~Bagdasarian\fnref{Berkeley}}
\author[Milano]{D.~Basilico}
\author[Milano]{G.~Bellini}
\author[PrincetonChemEng]{J.~Benziger}
\author[LNGS]{R.~Biondi}
\author[Milano]{D.~Bravo\fnref{Madrid}}
\author[Milano]{B.~Caccianiga}
\author[Princeton]{F.~Calaprice}
\author[Genova]{A.~Caminata}
\author[Virginia]{P.~Cavalcante\fnref{LNGSG}}
\author[Lomonosov]{A.~Chepurnov}
\author[Milano]{D.~D'Angelo}
\author[Genova]{S.~Davini}
\author[Peters]{A.~Derbin}
\author[LNGS]{A.~Di Giacinto}
\author[LNGS]{V.~Di Marcello}
\author[Princeton]{X.F.~Ding}
\author[Princeton]{A.~Di Ludovico} 
\author[Genova]{L.~Di Noto}
\author[Peters]{I.~Drachnev}
\author[Dubna,Milano]{A.~Formozov}
\author[APC]{D.~Franco}
\author[Princeton,GSSI]{C.~Galbiati}
\author[LNGS]{C.~Ghiano}
\author[Milano]{M.~Giammarchi}
\author[Princeton]{A.~Goretti\fnref{LNGSG}}
\author[Juelich,RWTH]{A.S.~G\"ottel}
\author[Lomonosov,Dubna]{M.~Gromov}
\author[Mainz]{D.~Guffanti}
\author[LNGS]{Aldo~Ianni}
\author[Princeton]{Andrea~Ianni}
\author[Krakow]{A.~Jany}
\author[Munchen]{D.~Jeschke}
\author[Kiev]{V.~Kobychev}
\author[London,Atomki]{G.~Korga}
\author[Juelich,RWTH]{S.~Kumaran}
\author[LNGS]{M.~Laubenstein}
\author[Kurchatov,Kurchatovb]{E.~Litvinovich}
\author[Milano]{P.~Lombardi}
\author[Peters]{I.~Lomskaya}
\author[Juelich,RWTH]{L.~Ludhova}
\author[Kurchatov]{G.~Lukyanchenko}
\author[Kurchatov]{L.~Lukyanchenko}
\author[Kurchatov,Kurchatovb]{I.~Machulin}
\author[Mainz]{J.~Martyn}
\author[Milano]{E.~Meroni}
\author[Dresda]{M.~Meyer}
\author[Milano]{L.~Miramonti}
\author[Krakow]{M.~Misiaszek}
\author[Peters]{V.~Muratova}
\author[Munchen]{B.~Neumair}
\author[Mainz]{M.~Nieslony}
\author[Kurchatov,Kurchatovb]{R.~Nugmanov}
\author[Munchen]{L.~Oberauer}
\author[Mainz]{V.~Orekhov}
\author[Perugia]{F.~Ortica}
\author[Genova]{M.~Pallavicini}
\author[Munchen]{L.~Papp}
\author[Juelich,RWTH]{L.~Pelicci}
\author[Juelich]{\"O.~Penek}
\author[Princeton]{L.~Pietrofaccia}
\author[Peters]{N.~Pilipenko}
\author[UMass]{A.~Pocar}
\author[Kurchatov]{G.~Raikov}
\author[LNGS]{M.T.~Ranalli}
\author[Milano]{G.~Ranucci}
\author[LNGS]{A.~Razeto}
\author[Milano]{A.~Re}
\author[Juelich,RWTH]{M.~Redchuk\fnref{Padova}}
\author[Perugia]{A.~Romani}
\author[LNGS]{N.~Rossi}
\author[Munchen]{S.~Sch\"onert}
\author[Peters]{D.~Semenov}
\author[Juelich]{G.~Settanta\fnref{ISPRA}}
\author[Kurchatov,Kurchatovb]{M.~Skorokhvatov}
\author[Juelich,RWTH]{A.~Singhal}
\author[Dubna]{O.~Smirnov}
\author[Dubna]{A.~Sotnikov}
\author[LNGS,Kurchatov]{Y.~Suvorov\fnref{Napoli}}
\author[LNGS]{R.~Tartaglia}
\author[Genova]{G.~Testera}
\author[Dresda]{J.~Thurn}
\author[Peters]{E.~Unzhakov}
\author[Dubna]{A.~Vishneva}
\author[Virginia]{R.B.~Vogelaar}
\author[Munchen]{F.~von~Feilitzsch}
\author[GSI,Juelich,RWTH]{A.~Wessel}
\author[Krakow]{M.~Wojcik}
\author[Hamburg]{B.~Wonsak}
\author[Mainz]{M.~Wurm}
\author[Genova]{S.~Zavatarelli}
\author[Dresda]{K.~Zuber}
\author[Krakow]{G.~Zuzel}

\fntext[Berkeley]{Present address: University of California, Berkeley, Department of Physics, CA 94720, Berkeley, USA}
\fntext[Napoli]{Present address: Dipartimento di Fisica, Universit\`a degli Studi Federico II e INFN, 80126 Napoli, Italy}
\fntext[Madrid]{Present address: Universidad Autónoma de Madrid, Ciudad Universitaria de Cantoblanco, 28049 Madrid, Spain}
\fntext[LNGSG]{Present address: INFN Laboratori Nazionali del Gran Sasso, 67010 Assergi (AQ), Italy}
\fntext[Padova]{Present address: Dipartimento di Fisica e Astronomia dell’Università di Padova and INFN Sezione di
Padova, Padova, Italy}
\fntext[ISPRA]{Present address: Istituto Superiore per la Protezione e la Ricerca Ambientale, 00144 Roma, Italy}
\address{\bf{The Borexino Collaboration}}
\address[APC]{APC, Universit\'e de Paris, CNRS, Astroparticule et Cosmologie, Paris F-75013, France}
\address[Dubna]{Joint Institute for Nuclear Research, 141980 Dubna, Russia}
\address[Genova]{Dipartimento di Fisica, Universit\`a degli Studi e INFN, 16146 Genova, Italy}
\address[Krakow]{M.~Smoluchowski Institute of Physics, Jagiellonian University, 30348 Krakow, Poland}
\address[Kiev]{Institute for Nuclear
Research of NAS Ukraine, 03028 Kyiv, Ukraine}
\address[Kurchatov]{National Research Centre Kurchatov Institute, 123182 Moscow, Russia}
\address[Kurchatovb]{ National Research Nuclear University MEPhI (Moscow Engineering Physics Institute), 115409 Moscow, Russia}
\address[LNGS]{INFN Laboratori Nazionali del Gran Sasso, 67010 Assergi (AQ), Italy}
\address[Milano]{Dipartimento di Fisica, Universit\`a degli Studi e INFN, 20133 Milano, Italy}
\address[Perugia]{Dipartimento di Chimica, Biologia e Biotecnologie, Universit\`a degli Studi e INFN, 06123 Perugia, Italy}
\address[Peters]{St. Petersburg Nuclear Physics Institute NRC Kurchatov Institute, 188350 Gatchina, Russia}
\address[Princeton]{Physics Department, Princeton University, Princeton, NJ 08544, USA}
\address[PrincetonChemEng]{Chemical Engineering Department, Princeton University, Princeton, NJ 08544, USA}
\address[UMass]{Amherst Center for Fundamental Interactions and Physics Department, UMass, Amherst, MA 01003, USA}
\address[Virginia]{Physics Department, Virginia Polytechnic Institute and State University, Blacksburg, VA 24061, USA}
\address[Munchen]{Physik-Department, Technische Universit\"at  M\"unchen, 85748 Garching, Germany}
\address[Lomonosov]{Lomonosov Moscow State University Skobeltsyn Institute of Nuclear Physics, 119234 Moscow, Russia}
\address[GSSI]{Gran Sasso Science Institute, 67100 L'Aquila, Italy}
\address[Dresda]{Department of Physics, Technische Universit\"at Dresden, 01062 Dresden, Germany}
\address[Mainz]{Institute of Physics and Cluster of Excellence PRISMA+, Johannes Gutenberg-Universit\"at Mainz, 55099 Mainz, Germany}
\address[GSI]{GSI Helmholtzzentrum f\"ur Schwerionenforschung, Planckstrasse 1, D-64291 Darmstadt, Germany}
\address[Juelich]{Institut f\"ur Kernphysik, Forschungszentrum J\"ulich, 52425 J\"ulich, Germany}
\address[RWTH]{III. Physikalisches Institut B, RWTH Aachen University, 52062 Aachen, Germany}
\address[London]{Department of Physics, School of Engineering, Physical and Mathematical Sciences, Royal Holloway, University of London, Egham, TW20 OEX, UK}
\address[London1]{Department of Physics and Astronomy, University College London, London, UK}
\address[Atomki]{Institute of Nuclear Research (Atomki), Debrecen, Hungary}
\address[Hamburg]{University of Hamburg, Institute of Experimental Physics, Luruper Chaussee 149, 22761 Hamburg, Germany}

\begin{abstract}
    We report the measurement of sub-MeV solar neutrinos through the use of their associated Cherenkov radiation, performed with the Borexino detector at the Laboratori Nazionali del Gran Sasso. The measurement is achieved using a novel technique that correlates individual photon hits of events to the known position of the Sun. In an energy window between 0.54\,MeV to 0.74\,MeV, selected using the dominant scintillation light, we have measured 10887$^{+2386}_{-2103} (\mathrm{stat.})\pm 947 (\mathrm{syst.})$ ($68\%$ confidence interval) solar neutrinos out of 19904 total events. This corresponds to a $^{7}$Be neutrino interaction rate of 51.6$^{+13.9}_{-12.5}$\,counts/(day$\cdot$\SI{100}{ton}), which is in agreement with the Standard Solar Model predictions and the previous spectroscopic results of  Borexino. The no-neutrino hypothesis can be excluded with $>$5$\sigma$\, confidence level. For the first time, we have demonstrated the possibility of utilizing the directional Cherenkov information for sub-MeV solar neutrinos, in a large-scale, high light yield liquid scintillator detector.
    This measurement provides an experimental proof of principle for future hybrid event reconstruction using both Cherenkov and scintillation signatures simultaneously.
\end{abstract}
     
    \end{frontmatter}
    \twocolumn 

{\it Introduction:}  Due to their tiny interaction cross-sections, neutrinos provide us with unique information about otherwise inaccessible locations such as the center of stars and other astronomical objects. Solar neutrinos are created as electron-flavored neutrinos in the nuclear fusion processes inside the Sun's core. They have a special place in neutrino physics as they not only aid us in understanding our Sun and what powers it~\cite{nature-phase2,CNO-nature,super-kamiokande_IV}, but also help us in unravelling various neutrino properties such as neutrino oscillations and the effect of neutrino-matter interactions~\cite{Homestake,GALLEX,SAGE,SNO_2001_PRL,SK_osc_2001,MSW_2004}. Moreover, neutrinos from the Sun enable us to put upper limits on the neutrino magnetic moment~\cite{Bx_magnetic} and the accumulation of dark matter~\cite{SK_DM} inside the Sun.

At present, there are two principal detector types used for the measurement of solar neutrinos: Water Cherenkov (WCh)~\cite{super-kamiokande_IV, SNO_water, SNO_LETA} and Liquid Scintillator (LS) detectors~\cite{detector-paper, KamLAND_solar, SNO_LS, JUNO}. They detect solar neutrinos via elastic scattering off electrons. The light produced by the recoil electrons is typically detected by photomultiplier tubes (PMTs). Both these detector types have their own sets of advantages and disadvantages.
In WCh detectors, the Cherenkov light emission from the recoil electrons enables the directional reconstruction of the final-state lepton, that is essential for background suppression and particle identification based on the Cherenkov ring morphology~\cite{super-kamiokande_IV, SK_reco_2019}. The main disadvantage of WCh detection is a relatively 
small light yield at MeV energies, resulting in a higher energy threshold as well as poorer resolution compared to LS detectors. Only charged particles that have a velocity faster than the speed of light in the medium can be detected. This is determined by the refractive index $n$ of the medium. For water with {\it n}$\approx$1.33, this results in a kinetic energy threshold of about \SI{0.25}{MeV} for electrons.
In practice, the effective low energy threshold is higher due to the presence of radioactive background and PMT dark noise. Since the amount of Cherenkov light emitted close to the threshold is small, this low light yield makes it challenging to trigger an event and perform vertex and direction reconstruction.
For example, taking into account the coverage, photon detection efficiency, and radioactive background, the lowest kinetic energy threshold for recoil electrons used in WCh detectors is $\sim$\SI{3.5}{MeV}, where only $\sim$30 photoelectron hits are detected on average~\cite{super-kamiokande_IV, SNO_LETA}.

On the other hand, large LS detectors have a relatively high light yield, and thus a higher energy resolution and a lower energy threshold, provided they have a sufficiently low level of residual radioactive contamination.
The recoil electrons from solar neutrinos excite the liquid scintillator molecules, which produce isotropic scintillation light. For Borexino this corresponds to 500 photoelectron hits at \SI{1}{MeV} deposited energy detected with 2000 PMTs. This is equivalent to a 5$\%$ energy resolution~\cite{BxCalibPaper} and an effective low energy threshold of $\sim$\SI{0.19}{MeV} for the spectroscopic analysis~\cite{phase2-nusol}.
The main disadvantage of LS detectors is that the emitted Cherenkov photons are sub-dominant to such a degree that an event-by-event direction reconstruction has not been possible yet.

There is an ongoing effort for the development of hybrid detectors that could combine the advantages of both detector types i.e. a low energy threshold, good energy resolution, and directional reconstruction using Cherenkov light~\cite{Theia}. 
This is motivated by the prospect of a rich physics program, ranging from the measurement of CNO solar neutrinos~\cite{CNO_directionality} to searches for neutrinoless double beta decay~\cite{NLDBD_dir} for which solar neutrinos are a background. Moreover, the added scintillation signal provides a means to reconstruct hadronic recoils in the final states of GeV neutrino interactions, most relevant for future long-baseline oscillation experiments~\cite{Theia}. 
The research and development activities for these hybrid detectors can be summarized into four categories:
(1) new target materials~\cite{WbLS,slowLS,QDLS}, (2) fast photo detectors for time separation of Cherenkov and scintillation light~\cite{LAPPD}, (3) spectral sorting with bandpass and dichroic filters~\cite{photon_sorting_1, photon_sorting_2}, and (4) new analysis techniques~\cite{LS_analysis_1, LS_analysis_2}.
Only relatively small-scale experiments have been run so far, or are planned to run in the near future, such as CHESS~\cite{CHESS_2}, ANNIE~\cite{ANNIE}, and FlatDot~\cite{FlatDot}. Contrariwise direction reconstruction with Cherenkov light in large-scale scintillator/hybrid neutrino detectors has only been studied with Monte-Carlo simulations~\cite{CNO_directionality, Theia, WbLS_MC_1}.

In this letter, we present a measurement of sub-MeV solar neutrinos in the Borexino detector, using their associated Cherenkov light. For this, a specific energy region of interest has been selected using the recoil electron scintillation light signal. The measurement is performed through a novel analysis technique called ``Correlated and Integrated Directionality'' (CID). This work provides an experimental proof of principle for the feasibility of using directional Cherenkov light in a monolithic, large-scale liquid scintillator detector. At the same time, the CID method also provides a robust, straightforward analysis technique that is readily applicable for other LS detectors like KamLAND~\cite{KamLAND_solar}, JUNO~\cite{JUNO}, and SNO+~\cite{SNO_LS}.

\begin{figure}[t!]
    \centering
    \includegraphics[width=0.49\textwidth]{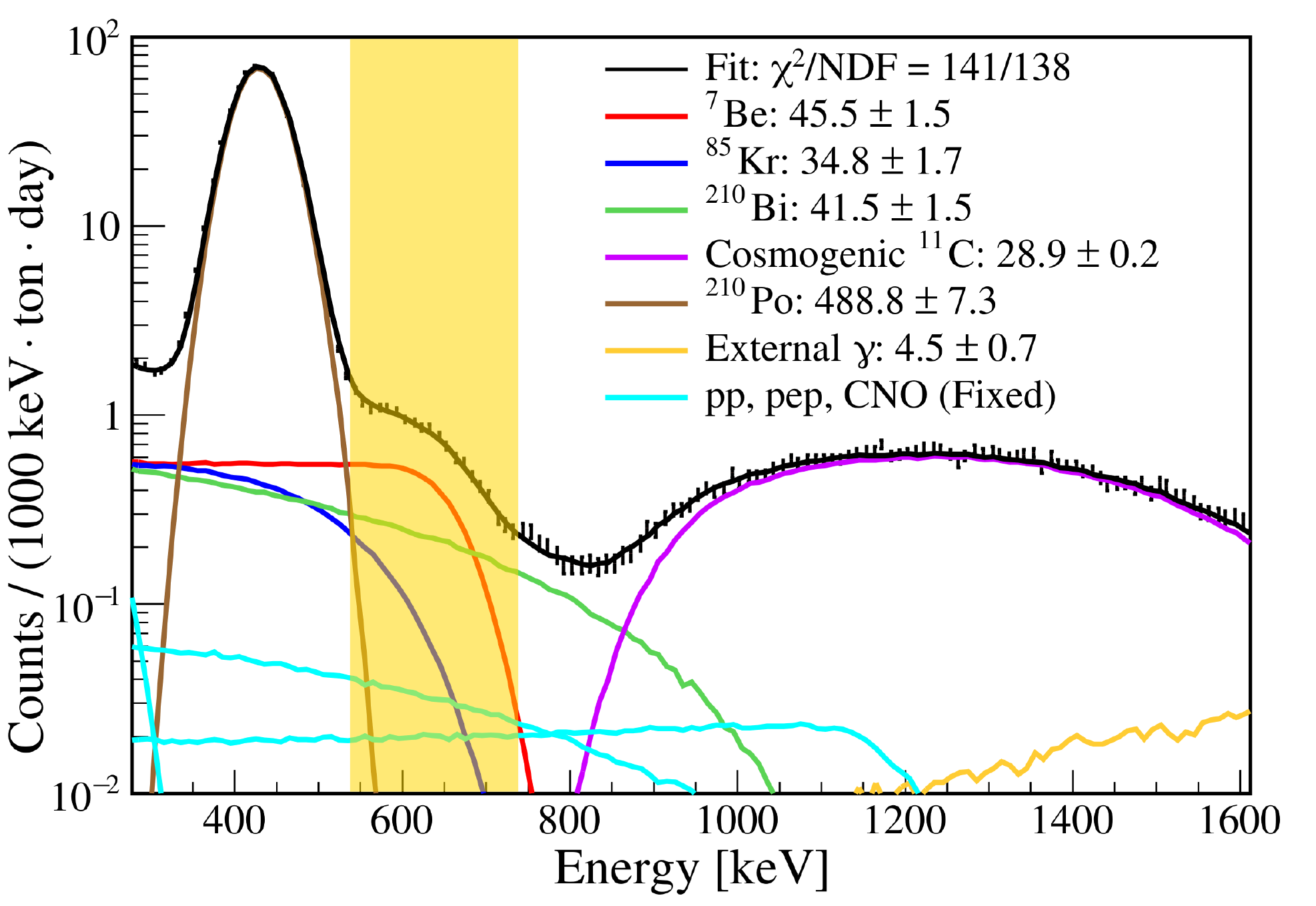}
    \caption{The energy spectrum of Phase-I data (black points) along with the spectral fit (black line) performed using the PDFs of the different solar neutrino components and backgrounds~\cite{phase1-nusol}. The energy Region of Interest (ROI) used for the directional analysis of Phase-I is shown as a shaded yellow area. }
    \label{fig:nusol-phase2-fit}
\end{figure}

{\it Borexino and solar neutrinos:} Borexino is a high light-yield LS detector whose main goal is the spectroscopic measurement of solar neutrinos~\cite{detector-paper}. The experiment is located at the Laboratori Nazionali del Gran Sasso (LNGS), Italy, at a depth of 3800 meters water equivalent to suppress the cosmic muon flux. The neutrino target at the detector center consists of $\sim$280\,tons of extremely radio-pure LS. It is composed of 1,2,4-trimethylbenzene (PC) solvent, mixed with 2,5-diphenyloxazole (PPO) as a fluor. The LS is contained in a thin nylon vessel, surrounded by two liquid buffer layers. The scintillator and buffers are contained in a stainless steel sphere, equipped with $\sim$2000 8-inch PMTs. This entire setup is enclosed in a water tank mounted with $\sim$200 PMTs serving as a muon veto. 
The experimental dataset is divided into three main Phases: Phase-I (May 2007\textendash May 2010), Phase-II (December 2011\textendash May 2016), and Phase-III (July 2016\textendash October 2021). 
Borexino's ability to measure solar neutrinos through spectral fits has already been well-demonstrated in the past years through the complete spectroscopy of \emph{pp}-chain neutrinos~\cite{nature-phase2, Be7-paper, pep}
and the first direct detection of CNO neutrinos~\cite{CNO-nature}. 
The directionality measurement reported here is performed on individual photon hits of the events selected in a specific energy Region of Interest (ROI) to increase the signal-to-background ratio. This facilitates the statistical separation of solar neutrinos and the intrinsic, isotropic radioactive background in the LS. The signal consists of electrons scattering off 0.862\,MeV mono-energetic $^{7}$Be solar neutrinos ($\sim$90\% of signal), 1.44\,MeV mono-energetic \emph{pep} solar neutrinos, and CNO solar neutrinos with an end point at 1.74\,MeV. Solar neutrinos are named corresponding to the reaction process in which they are created. Since solar neutrinos are detected via their elastic scattering off electrons in the LS, even mono-energetic neutrinos show a continuous spectrum, characterized by a Compton-like edge. In the case of $^{7}$Be solar neutrinos, this edge is at around 0.66\,MeV (see Figure~\ref{fig:nusol-phase2-fit}). The main backgrounds for this analysis include intrinsic radioactive $\beta^{-}$ emitters, namely $^{210}$Bi ($Q$=1.162\,MeV) and $^{85}$Kr ($Q$=0.687\,MeV). 

{\it Cherenkov light in Borexino:} Due to the wavelength dependence ($\lambda^{-2}$) of the Cherenkov spectrum, most photons are produced in the Ultra-Violet (UV) region. These Cherenkov photons are absorbed by the fluor PPO and are subsequently re-emitted as scintillation light. In Borexino, both Cherenkov and scintillation light spectra are detected above a wavelength of 370\,nm, where the PPO absorption becomes negligible. The spectra and velocity of Cherenkov and scintillation light in the LS depend on the wavelength-dependent refractive index (Figure 9 in~\cite{Borex-mc}).  Considering $n\approx$1.55 at 400\,nm, Cherenkov light is produced when the kinetic energy of the recoil electron from a solar neutrino exceeds 0.16\,MeV.
Cherenkov light is emitted at picosecond time-scale while the fastest scintillation light component from the LS has an emission time constant at the nanosecond level.
For this reason, the observed ratio of Cherenkov photons in the first few nanoseconds of a recorded event is considerably higher than for the entire event time. The detected PMT hit time distributions are broadened by various optical processes in the LS, the Transit Time Spread (TTS) of the PMTs, jitter of the electronics, and precision of PMT time calibration. All the above-mentioned effects, as well as the multiple scattering of the recoil electrons in the LS, are taken into account in the customized Geant-4 based Monte-Carlo (MC) simulation developed for Borexino~\cite{Borex-mc}. The parameters of the MC have been tuned to reproduce the signals of radioactive sources used in the calibration campaign~\cite{BxCalibPaper}. 

\begin{figure}[t]
    \centering
    \includegraphics[width=0.49\textwidth]{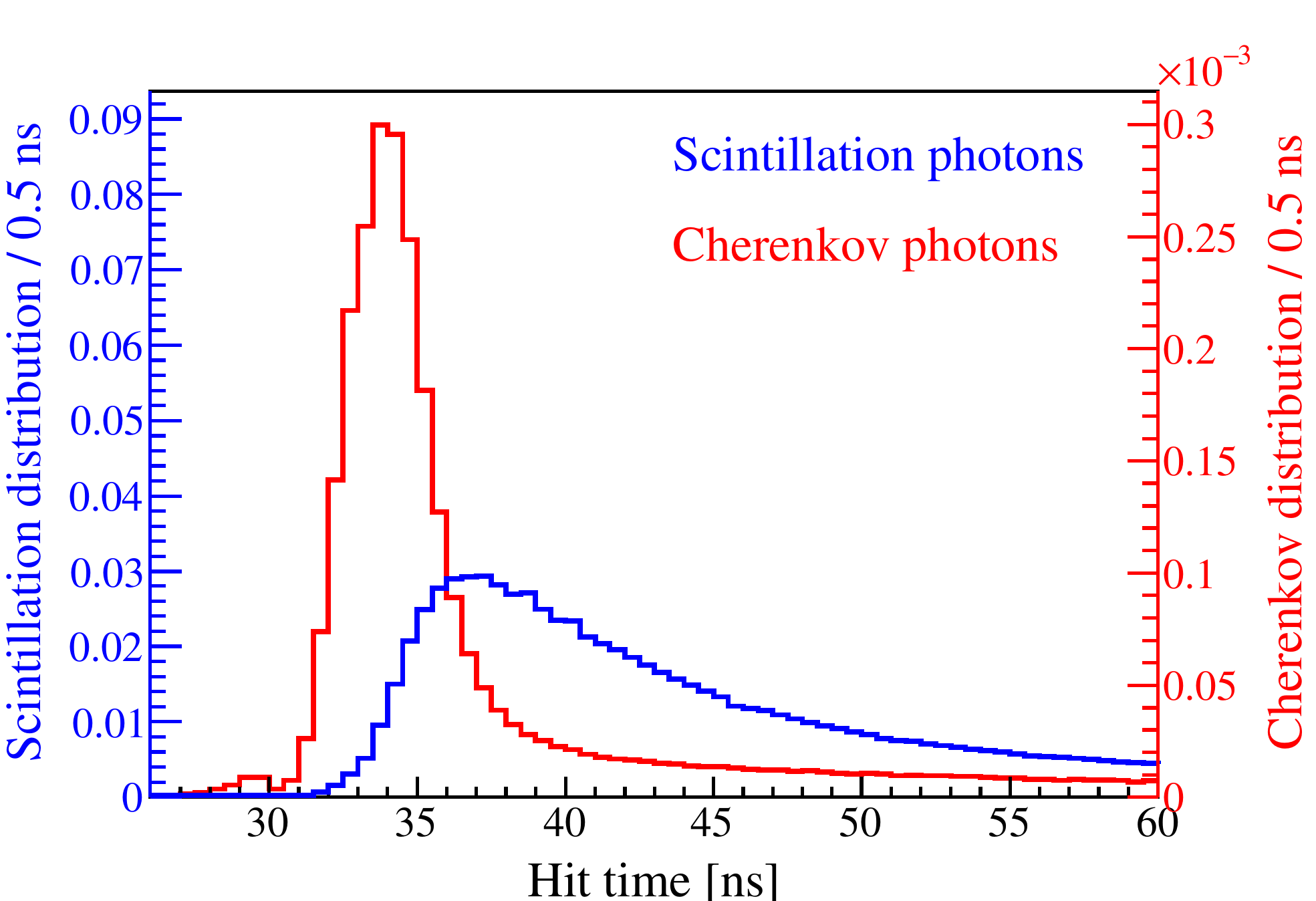}
    \caption{Time-of-flight corrected hit times of $^{7}$Be solar neutrino recoil electrons (\SI{0.54}{MeV} to \SI{0.74}{MeV}) as obtained from the Borexino MC. The left y-axis shows the scintillation light (blue), where the area is normalized to 1. The Cherenkov light (red) is shown on the right y-axis and the area is normalized to the number of Cherenkov hits relative to scintillation ($\sim$0.4\%).  The scintillation light profile also includes photons that have been produced in the Cherenkov process, but have been absorbed and re-emitted by the LS.}
    \label{fig:timepdf}
\end{figure}

The distinct time behaviors of Cherenkov and scintillation photons are illustrated in Figure~\ref{fig:timepdf}. It shows the time-of-flight corrected PMT hit time distributions obtained from MC for $^{7}$Be solar neutrino recoil electrons in the ROI.
In the chosen energy ROI, on average $\sim$270 PMT hits (normalized to 2000 live PMTs) are detected per event and MC predicts that only $\sim$1 PMT hit per event is caused by Cherenkov light. This highly disfavors an event-by-event directional reconstruction in Borexino.

\begin{figure}[t]
    \centering
    \includegraphics[width=0.47\textwidth]{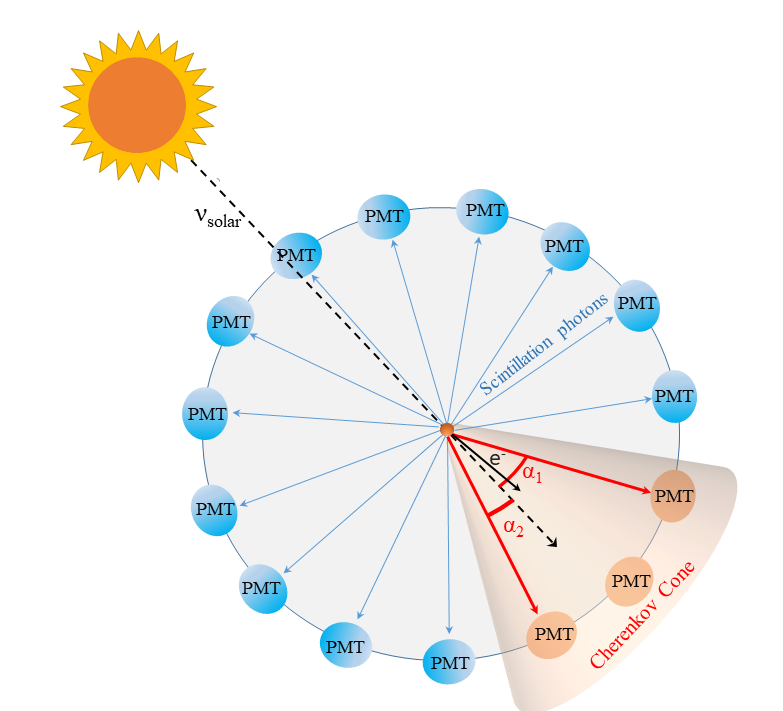}
    \caption{Angular correlation of photon hits expressed in terms of the directional angle ${\alpha}$ given by the reconstructed vertex of the solar neutrino event and the position of the Sun. An electron recoiling off a solar neutrino  in the detector produces isotropic scintillation light (blue arrows) uncorrelated to the Sun, as well as a Cherenkov cone (orange arrows) in the solar direction. In this example, the first two event hits are both Cherenkov photons and their respective directional angles are $\alpha_{1}$ and $\alpha_{2}$. The direction of Cherenkov and scintillation photons of radioactive background events are not correlated to the position of the Sun.}
    \label{fig:cos_alpha_sketch}
\end{figure}

\begin{figure*}[t!]
    \centering
    \subfigure[]{\includegraphics[width=0.505\textwidth]{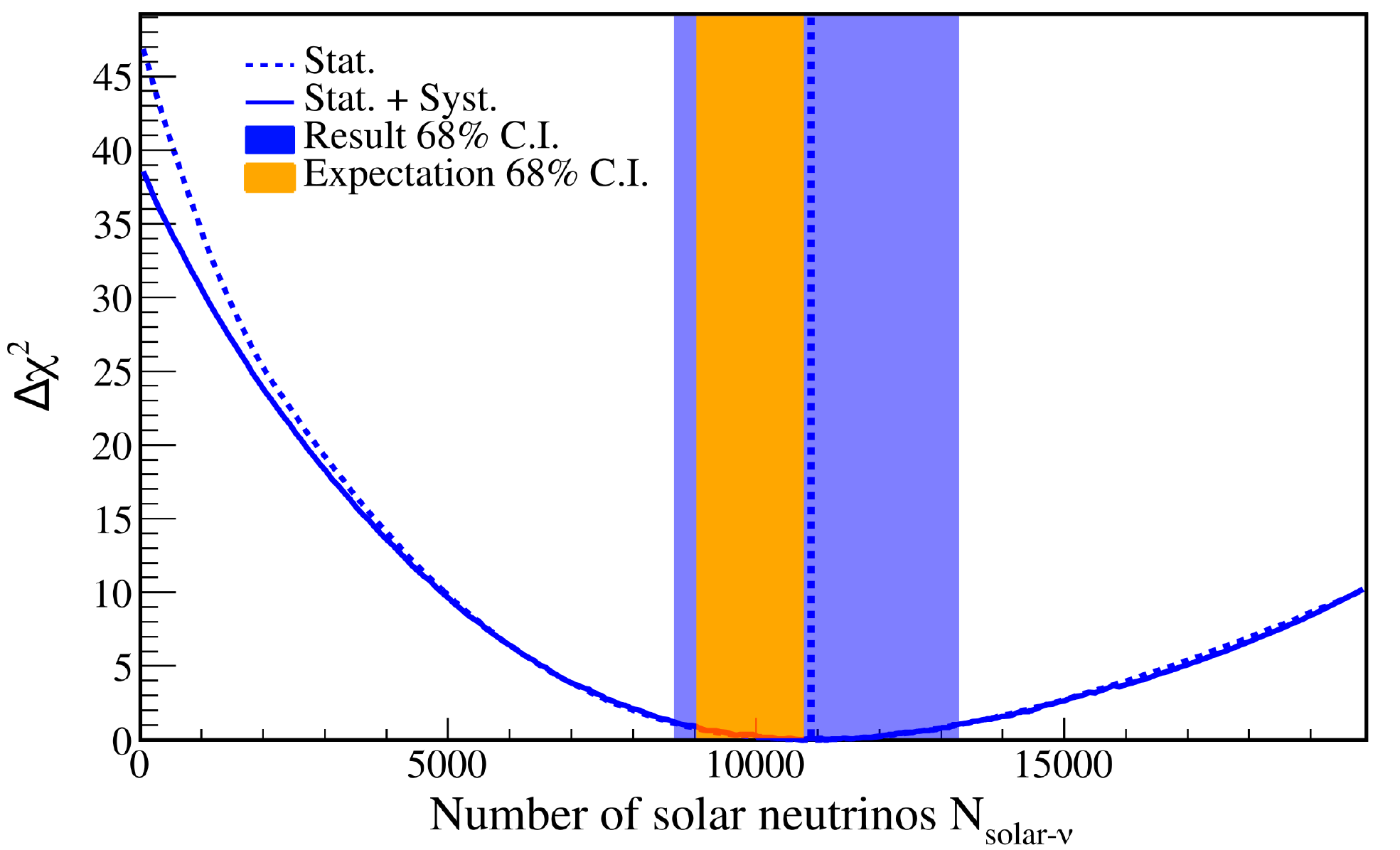}\label{fig:chi2_sigtotot}}
    \subfigure[]{\includegraphics[width=0.48\textwidth]{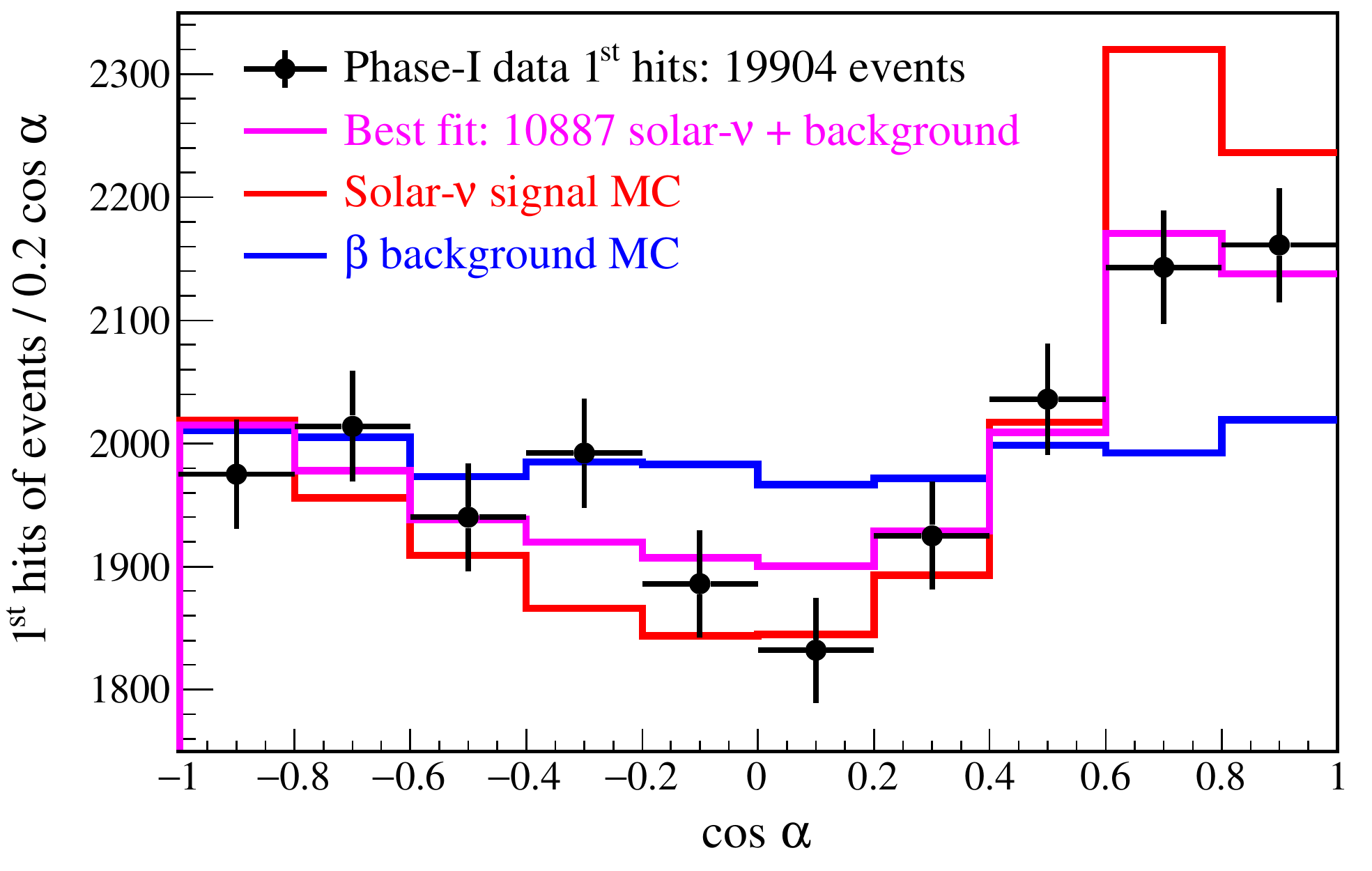}    \label{fig:bestfit_dataMC_1}}
    \caption{(a) The sum of the $\Delta\chi^{2}$ profiles of the 1$^{\mathrm{st}}$ and 2$^{\mathrm{nd}}$ hits from the fit as a function of the observed number of solar neutrinos $N_{\mathrm{solar}-\nu}$, with (blue solid curve) and without (blue dotted curve) the systematic uncertainty. The no-neutrino signal hypothesis $N_{\mathrm{solar}-\nu}$\,=\,0, can be rejected with $\Delta\chi^{2}>25$,  $>5\sigma$ C.L.  The 68\%\,C.I. (blue shaded band) from the $\Delta\chi^{2}$ profile gives $N_{\mathrm{solar}-\nu}$ = 10887$^{+2386}_{-2103} (\mathrm{stat.})\pm 947 (\mathrm{syst.})$. The best fit gives a $\chi^{2}/ndf$ = 124.6/117. The 68\%\,C.I. of the expected solar neutrino signal based on the Standard Solar Model (SSM) predictions~\cite{CNOsens} is shown as an orange band. (b) The $\cos{\alpha}$ distributions of the 1$^{\mathrm{st}}$ hits of all the selected events (black points) compared with the best fit curve (magenta) of the resulting number of solar neutrinos plus background. The MC PDFs of pure neutrino signal (red) and $\beta$ background (blue) used in the fit are shown as well, normalized to the same 19904 events. For illustration purposes, the histograms are shown with 10 bins instead of the 60 bins as used in the final fit. }
    \label{fig:results}
\end{figure*}

{\it Correlated and Integrated Directionality (CID):} 
In the CID technique, we correlate the detected PMT-hit pattern of a selected event to the well-known position of the Sun, and then integrate it for a large number of events. This results in an angular distribution between the hit PMTs and the solar direction which is illustrated in Figure~\ref{fig:cos_alpha_sketch}. Due to the event kinematics, the angular distribution of recoil electrons is centered around the direction of the incident solar neutrinos. The Cherenkov light is produced almost instantaneously and carries the directional information of the recoil electron, and thus approximately that of the solar neutrino. The dominant scintillation light is emitted isotropically and has no correlation to the Sun's position. Since events are detected in real time, the position of the Sun is well-known for each event. The directional angle $\alpha$ is defined for each PMT hit as the angle between the known solar direction and the photon direction, given by the reconstructed event vertex of the recoil electron and the hit PMT position. Given the energy ROI and the refractive index of the LS, the CID distribution of solar neutrinos is expected to have a signature Cherenkov peak at $\cos{\alpha}$\,$\sim$0.7 (see Figure~\ref{fig:bestfit_dataMC_1}). Since the Cherenkov and scintillation light from the radioactive background inside Borexino has no correlation to the Sun's position, it instead produces a flat CID $\cos{\alpha}$ distribution. These CID distributions of solar neutrinos and background events can be disentangled by fitting MC-generated Probability Density Functions (PDFs) for the directional signal and flat background to a high-statistics data set.  Note that the CID method on its own is largely insensitive to different kinds of solar neutrinos like $^{7}$Be, \emph{pep}, and CNO in the chosen ROI.

{\it Analysis methods: } The principal dataset of this analysis consists of Phase-I. The data selection for this measurement follows a procedure similar to the standard low-energy solar neutrino analyses of Borexino~\cite{phase1-nusol, phase2-nusol}. However, the events are selected in a smaller energy ROI between 0.54-0.74\,MeV (Figure~\ref{fig:nusol-phase2-fit}). In this ROI, external gammas originating from outside the LS are still negligible until a radius $r < 3.3$\,m, allowing us to use an enlarged fiducial volume of 132\,t. In addition, pulse shape discrimination~\cite{Geo-longpaper} is applied to remove $\alpha$ background from radioactive $^{210}$Po ($Q$=5.304\,MeV) in the scintillator. Its quenched energy spectrum (0.28-0.63\,MeV electron-equivalent) falls partially inside the ROI. In the chosen ROI, the overall ratio of Cherenkov to scintillation photons expected from MC is only $\sim$0.4\%. 
Unlike the usual event-based analyses of Borexino, the CID method is performed on the individual photon hits. As a consequence, we have adopted the \emph{``N\textsuperscript{th}-hit method''}: The times of the photon hits of each selected event are corrected with their time-of-flight between the reconstructed event vertex and the PMT that detected the hit. They are then are sorted in time with respect to the reconstructed start time of the event. The analysis is restricted to the 1\textsuperscript{st} and 2\textsuperscript{nd} hits as they were found to have the highest Cherenkov ratio, capable of distinguishing the directional differences between signal and background in MC. For both 1\textsuperscript{st} and 2\textsuperscript{nd} hits separately, CID $\cos{\alpha}$ distributions are produced by summing over the hits of all selected events. 

The number of solar neutrinos $N_{\mathrm{solar}-\nu}$ is extracted through a $\chi^{2}$-fit of the CID $\cos{\alpha}$ data spectrum to the MC PDFs of the solar neutrino signal and $\beta^{-}$ background. The test statistics is defined as follows:
\begin{align}\label{eq:nusol-fit}
\nonumber
    &\chi^{2}(N_{\mathrm{solar}-\nu}) = \\ \nonumber
    & \sum \limits_{n=1}^{N}\sum\limits_{i=1}^{I}
    \left(
    \frac{\bigg((\cos{\alpha})^{D}_{n, i} - (\cos{\alpha})^{M}_{n, i}\big(N_{\mathrm{solar}-\nu}, \Delta r_{\mathrm{dir}}, gv_{\mathrm{ch}}^{\mathrm{corr}}\big)\bigg)^{2}}{(\sigma^{D}_{n, i})^{2} + (\sigma^{M}_{n, i})^{2}} 
    \right.\\ 
    &+ \left.\frac{(gv_{\mathrm{ch}}^{\mathrm{corr}} - \SI{0.108}{\nano\second\per\meter})^2}{(\SI{0.039}{\nano\second\per\meter})^2} 
    \right),
\end{align}
with the hit index $n=1, 2$ and the angular index $i$ from 1 to the total number of bins $I=60$ in the range -1\,$< \cos{\alpha} <$\,+1. $(\cos{\alpha})^{D}_{n, i}$ and $(\cos{\alpha})^{M}_{n, i}$ are the $\cos{\alpha}$ values for the i\textsuperscript{th} bin of the n\textsuperscript{th} hit of data and MC, respectively, and, $\sigma^{D}_{n,i}$ and $\sigma^{M}_{n, i}$ are their corresponding statistical errors. The parameters $\Delta r_{\mathrm{dir}}$ and $gv_{\text{ch}}^{\mathrm{corr}}$ are sufficient to parameterize the differences between data and MC. Their values are small and have no impact on the event-based algorithms of Borexino.
The parameter $\Delta r_{\mathrm{dir}}$ takes into account a systematic shift in the reconstructed vertex position of the recoil electron with respect to the initial electron direction, which is correlated to the solar neutrino direction. It is not visible on an event-by-event basis, but is observed in MC and has an influence on the summed CID $\cos{\alpha}$ distribution of solar neutrinos. The value of $\Delta r_{\mathrm{dir}}$ in data cannot be determined by Borexino calibrations and hence is left free to vary in the fit.
The parameter $gv_{\text{ch}}^{\mathrm{corr}}$ is an effective correction of the group velocity for Cherenkov photons relative to scintillation photons and is treated as a nuisance parameter with a Gaussian pull-term in the fit, based on gamma calibration data as explained below. This correction term reflects the remaining uncertainty on the effective Cherenkov wavelength spectrum and the wavelength-dependent refractive index implemented in the detector MC. These can affect the relative group velocities of scintillation and Cherenkov photons and thus change their effective ratio for the first few photon hits used in the analysis.

We have performed a calibration of the effective Cherenkov group velocity correction using the available gamma calibration sources from Borexino's 2009 calibration campaign~\cite{BxCalibPaper}. Due to the known positions of the gamma sources from CCD cameras, the initial directions of the gamma rays can be reconstructed. The Cherenkov photons from the Compton-scattered electrons can be correlated to the reconstructed gamma direction similar to the CID analysis. This calibration results in a value of $gv_{\text{ch}}^{\mathrm{corr}}$ = (0.108 $\pm$ 0.039)\,ns\,m$^{-1}$. Since the calibration has been performed at the end of Phase-I data-taking and sub-nanosecond stability of effective PMT timing cannot be guaranteed for long time periods, this measurement is considered valid only for Phase-I. Note that Cherenkov photons from background PDFs are not influenced by $\Delta r_{\mathrm{dir}}$ and $gv_{\text{ch}}^{\mathrm{corr}}$ as the direction of both Cherenkov and scintillation light are uncorrelated to the position of the Sun. Both $\Delta r_{\mathrm{dir}}$ and $gv_{\text{ch}}^{\mathrm{corr}}$ are explained in more detail in~\cite{arxivprd}. Note that in future LS detectors, the uncertainties arising from both parameters might be substantially reduced by the deployment of a dedicated electron Cherenkov source.

Further systematic effects have been investigated: the choice of the histogram binning, the number of early PMT N\textsuperscript{th} hits, and the influence of the number and distribution of active PMTs. In total, they contribute a relative systematic uncertainty of $8.7\%$ on the $N_{\mathrm{solar}-\nu}$ result.
Other sources of uncertainty, such as the effect of a non-uniform distribution of radioactive background in the fiducial volume have been studied, but found to be negligible (more details in~\cite{arxivprd}).

{\it Results:} Figure~\ref{fig:chi2_sigtotot} shows the $\Delta\chi^{2}$ between data and the best fit as a function of the number of solar neutrino events $N_{\mathrm{solar}-\nu}$.
We show the $\Delta\chi^{2}$ profile both with and without the systematic uncertainty of $8.7\%$.
The agreement between the best fit and data is given by $\chi^{2}/ndf$ = 124.6/117 ($p\mathrm{-value}=0.30$), based on the histograms of the first two PMT hits of all events, with 60 bins each. The resulting best fit for the number of solar neutrino events is $N_{\mathrm{solar}-\nu}$ = 10887$^{+2386}_{-2103} (\mathrm{stat.})\pm 947 (\mathrm{syst.})$ ($68\%$ C.I.) out of the 19904 selected events, and consists of $^{7}$Be, \emph{pep}, and CNO neutrinos. Figure~\ref{fig:chi2_sigtotot} also shows the $68\%$ confidence intervals for the measured $N_{\mathrm{solar}-\nu}$ and the expected value according to the Standard Solar Model (SSM)~\cite{Vinyoles:2016djt, CNOsens}, represented by the blue and orange bands, respectively.
The SSM has varied predictions on the so-called ``metallicity'', i.e. the amount of metals heavier than $^{4}$He in the Sun. Since this affects the expected number of solar neutrinos in Borexino, the difference between the low and high metallicity model predictions are included as a systematic uncertainty. The measured number of solar neutrinos is well in agreement with the SSM expectation of \emph{N$_{\mathrm{SSM}}$} = 10187$_{-1127}^{+541}$. The background-only hypothesis can be excluded with $\Delta\chi^2$\,$>$\,25, which corresponds to a $>$5$\sigma$ detection of sub-MeV solar neutrinos using the CID method.
Figure~\ref{fig:bestfit_dataMC_1} shows the measured $\cos{\alpha}$ distribution for the first PMT hits of the data events together with the best fit and the neutrino-only and background-only $\cos{\alpha}$ distributions. For illustration purposes, this is shown for 10 bins instead of the 60 bins used in the final fit.

Using the CID measurement of $N_{\mathrm{solar}-\nu}$, the $^{7}$Be interaction rate R($^{7}$Be)$_{\mathrm{CID}}$ has been calculated as R($^{7}\mathrm{Be})_{\mathrm{CID}}=51.6^{+13.9}_{-12.5}$\,cpd/\SI{100}{t} for the full $^{7}$Be neutrino energy (0.384\,MeV and 0.862\,MeV monoenergetic lines). For this the \emph{pep} and CNO neutrino rates have been fixed to their SSM predictions~\cite{Vinyoles:2016djt}. The small errors arising from \emph{pep} and CNO neutrino predictions are included in the systematic uncertainty~\cite{arxivprd}.
This $^{7}$Be rate is well in agreement with the results of the Phase-I spectroscopy R($^{7}$Be) = 47.9$\pm$2.3~cpd/\SI{100}{t} where the \emph{pep} and CNO neutrino rates were fixed to their SSM predictions as well~\cite{phase1-nusol}.

{\it Conclusions:}
For the first time, we have measured sub-MeV solar neutrinos using their directional Cherenkov light, through the CID method in a traditional, large-scale LS detector. While this measurement on its own features relatively large statistical and systematic uncertainties, it still provides experimental proof that the directional information of Cherenkov light is accessible even for sub-MeV neutrinos in a high light-yield LS medium.
In future analyses the CID measurement could be combined with a standard spectral fit, and thus help to disentangle neutrino signal and backgrounds. This might be interesting especially if there is a degeneracy of signal and background energy spectra, as is the case for the $^{210}$Bi background and CNO solar neutrinos~\cite{CNO-nature} in Borexino.

{\it Acknowledgements:}
We acknowledge the generous hospitality and support of the Laboratori Nazionali del Gran Sasso (Italy). The Borexino program is made possible by funding from Istituto Nazionale di Fisica Nucleare (INFN) (Italy), National Science Foundation (NSF) (USA), Deutsche Forschungsgemeinschaft (DFG), Cluster of Excellence PRISMA+ (Project ID 39083149), and recruitment initiative of Helmholtz-Gemeinschaft (HGF) (Germany), Russian Foundation for Basic Research (RFBR) (Grants No. 19-02-00097A) and Russian Science Foundation (RSF) (Grant No. 21-12-00063) (Russia), and Narodowe Centrum Nauki (NCN) (Grant No. UMO 2017/26/M/ST2/00915) (Poland). We gratefully acknowledge the computing services of Bologna INFN-CNAF data centre and U-Lite Computing Center and Network Service at LNGS (Italy).

% \bibliographystyle{apsrev4-1}
% \bibliography{main.bib}
\input{main.bbl}
\end{document}

%% file: main.bbl
%merlin.mbs apsrev4-1.bst 2010-07-25 4.21a (PWD, AO, DPC) hacked
%Control: key (0)
%Control: author (72) initials jnrlst
%Control: editor formatted (1) identically to author
%Control: production of article title (-1) disabled
%Control: page (0) single
%Control: year (1) truncated
%Control: production of eprint (0) enabled
%